\documentclass[twoside,fleqn]{article}
\usepackage{graphicx}
\usepackage{espcrc2}

\def\rp{$R_p \hspace{-1em}/\;\:$ }
\def\Table#1{Table~\ref{#1}}
\def\bold#1{\setbox0=\hbox{$#1$} 
     \kern-.025em\copy0\kern-\wd0 
     \kern.05em\copy0\kern-\wd0 
     \kern-.025em\raise.0433em\box0 }
\def\vev#1{\left\langle #1\right\rangle}
\def\beq{\begin{equation}}
\def\eeq{\end{equation}}
\def\Slash#1{#1\!\!\!\! /}
\def\beqa{\begin{eqnarray}}
\def\eeqa{\end{eqnarray}}
\def\vb#1{\vbox to #1 pt{}}
\def\nn{\nonumber}
\def\ds{\displaystyle}
\def\21{$SU(2) \otimes U(1)$}
\def\cimaum#1{\hbox{\raise .5ex\hbox{$#1$}}}

\def\ifmath#1{\relax\ifmmode #1\else $#1$\fi}
\def\half{\ifmath{{\textstyle{1 \over 2}}}}
\def\quarter{\ifmath{{\textstyle{1 \over 4}}}}
\def\Eq#1{{Eq. (\ref{#1})}}
\def\Fig#1{{Fig. (\ref{#1})}}

\def\npb#1#2#3{{\it Nucl.\ Phys.\ }{\bf B #1} (#2) #3}
\def\plb#1#2#3{{\it Phys.\ Lett.\ }{\bf B #1} (#2) #3}
\def\prd#1#2#3{{\it Phys.\ Rev.\ }{\bf D #1} (#2) #3}
\def\prep#1#2#3{{\it Phys.\ Rep.\ }{\bf #1} (#2) #3}
\def\prl#1#2#3{{\it Phys.\ Rev.\ Lett.\ }{\bf #1} (#2) #3}

\def\jetpl#1#2#3{{\it Sov.\ Phys.\ JETP Lett.\ }{\bf #1} (#2) #3}
\def\rnc#1#2#3{{\it Riv. Nuovo Cimento }{\bf #1} (#2) #3}

\def\hepph#1{{\tt hep-ph/#1}}

\newcommand {\ignore}[1]{}

\title{Neutrino Masses from Broken R-Parity}

\author{J. C. Rom\~ao\address{Instituto Superior T\'ecnico, 
Departamento de F\'{\i}sica\\
A. Rovisco Pais 1, 1049-001 Lisboa, Portugal}
        \thanks{This work was supported by the TMR network grant
ERBFMRXCT960090 of the European Union.}}

\begin{document}

\begin{abstract}
We review models where R--parity is broken, either spontaneously 
or explicitly. In this last case we consider the situation where
R--parity is broken via bilinear terms in the superpotential. We show that
although at tree level only one neutrino gets mass, at one--loop level
all three neutrinos became massive. We study the conditions under
which bimaximal mixing can be achieved and show that the masses can be
in the correct ranges needed for solving the atmospheric and solar neutrino
problems. 

\end{abstract}

\maketitle

\section{Introduction}

In the past most discussions of supersymmetric (SUSY)\cite{SUSY,MSSM}
phenomenology assumed R--parity ($R_P$) conservation where,
\beq
R_P=(-1)^{2J +3B +L}
\eeq
This implies that SUSY particles are pair produced, 
every SUSY particle decays into another SUSY particle and that
there is a {\it LSP} that it is stable.
But this is just an {\it ad hoc} assumption without a deep
justification. 
In this talk we will review how $R_P$ can be broken, either
spontaneously or explicitly, and discuss the most important features
of these models \cite{brasil98}. 
We will also describe recent results \cite{numass} on
one--loop generated masses and mixings in the context of a model that is a
minimal extension of the minimal extension of the MSSM--GUT
\cite{epsrad} in which $R_P$ Violation (RPV) is introduced via a
bilinear term in the MSSM superpotential \cite{e3others,chaHiggsEps}.

\section{Spontaneously Broken R-Parity}

\subsection{The Original Proposal}

In the original proposal~\cite{OriginalSBRP} the content was just the 
MSSM and the breaking was induced by
\beq
\vev{\tilde{\nu}_{\tau}} = v_L
\eeq
The problem with this model was that the Majoron $J$ coupled to $Z^0$ with 
gauge strength and therefore the decay
$Z^0 \rightarrow \rho_L J$ contributed to the invisible $Z$ width the
equivalent of half a (light) neutrino family. After LEP I this was
excluded.

\subsection{A Viable Model for SBRP}

The way to avoid the previous difficulty 
is to enlarge the model and make $J$ mostly out of {\it
isosinglets}. This was proposed by Masiero and Valle~\cite{MV}. The
content is the MSSM plus a few Isosinglet Superfields that carry
lepton number,
\beq
\nu^c_i\equiv (1,0,-1)\ ; \ S_i \equiv(1,0,1) \ ; \
\Phi\equiv (1,0,0)
\eeq
The model is defined by the superpotential \cite{MV,SBRpV},
\beqa
W&=&h_u u^c Q H_u + h_d d^c Q H_d + h_e e^c L H_d \cr
\vb{18}
&&+(h_0 H_u H_d - \mu^2 ) \Phi \cr
\vb{18}
&&+ h_{\nu} \nu^c L H_u + h \Phi \nu^c S \nn
\eeqa
where the lepton number assignments are shown in \Table{table0}.
\begin{table}
\caption{Lepton number assignments.}
\begin{tabular}{lccccc} \hline
Field & $L$ & $e^c$ & $\nu^c$ & $S$ & others  \\ 
Lepton \# & $1$ & $-1$ & $-1$& $1$ & $0$ \\ \hline
\end{tabular}
\label{table0}
\end{table}
The spontaneous breaking of R parity
and lepton number is driven by \cite{SBRpV}
\beq
v_R = \vev {\tilde{\nu}_{R\tau}} \quad
v_S = \vev {\tilde{S}_{\tau}} \quad
v_L = \vev {\tilde{\nu}_{\tau}}
\eeq
The electroweak breaking and fermion masses arise from
\beq
\vev {H_u} = v_u ~~~~~
\vev {H_d} = v_d
\eeq
with $v^2 = v_u^2 + v_d^2$ fixed by the W mass.
The Majoron is given by the imaginary part of 
\beq
\frac{v_L^2}{V v^2} (v_u H_u - v_d H_d) +
              \frac{v_L}{V} \tilde{\nu_{\tau}} -
              \frac{v_R}{V} \tilde{{\nu^c}_{\tau}} +
              \frac{v_S}{V} \tilde{S_{\tau}}
\eeq
where $V = \sqrt{v_R^2 + v_S^2}$. 
Since the Majoron
is mainly an \21 singlet it does not contribute to the
invisible $Z^0$ decay width.

\subsection{Some Results on SBRP}

The SBRP model has been extensively studied. The implications for
accelerator \cite{SBRPAcc} and non--accelerator \cite{SBRPNonAcc}
physics have been presented  before and we will not discuss them here
\cite{brasil98}. As in this talk we are concerned with the neutrino
properties in the
context of $R_P$ models we will only review here the neutrino
results. 

\begin{itemize}

\item
{\it Neutrinos have mass}

Neutrinos are massless at Lagrangian level but get mass from the
mixing with neutralinos\cite{paulo,npb}. In the SBRP model it is
possible to have non zero masses for two neutrinos \cite{npb}.

\item
{\it Neutrinos mix}

The mixing is related to the the 
coupling matrix $h_{\nu_{ij}}$. This matrix  has to be non diagonal in
generation space to allow
\beq
\nu_{\tau} \rightarrow \nu_{\mu} + J 
\eeq
and therefore evading~\cite{npb} the {\it Critical Density Argument} against
$\nu's$ in the MeV range.

\item
{\it Avoiding BBN constraints on the $m_{\nu_{\tau}}$}

In the {\it SM} BBN arguments~\cite{bbnothers} rule out $\nu_{\tau}$ 
masses in the range
\beq
0.5\ MeV < m_{\nu_{\tau}} < 35 MeV
\eeq
We have shown~\cite{bbnpaper} that {\it SBRP} models can evade that 
constraint due to new annihilation channels
\beq
\nu_{\tau} \nu_{\tau} \rightarrow J J 
\eeq

\end{itemize}

\section{Explicitly Broken R-Parity}

The most general superpotential $W$ with the particle content
of the MSSM is given by \cite{e3others,chaHiggsEps}

\beq
W= W_{MSSM} + W_{\Slash{R}}
\eeq
where 
\begin{eqnarray}
W_{MSSM}&=&\varepsilon_{ab}\left[
 h_U^{ij}\widehat Q_i^a\widehat U_j\widehat H_u^b
+h_D^{ij}\widehat Q_i^b\widehat D_j\widehat H_d^a\right.\cr
&&\cr
&&\left. \hskip 0.5cm
+h_E^{ij}\widehat L_i^b\widehat R_j\widehat H_d^a 
 -\mu\widehat H_d^a\widehat H_u^b \right]
\label{WMSSM}
\end{eqnarray}
and 
\begin{eqnarray}
W_{\Slash{R}}&=&\varepsilon_{ab}\left[
 \lambda_{ijk}\widehat L_i^a\widehat L_j^b\widehat R_k
+\lambda^{'}_{ijk}\widehat D_i \widehat L_j^a\widehat Q_k^b \right]\cr
&&\cr
&& 
+\lambda^{''}_{ijk} \widehat D_i \widehat D_j\widehat U_k 
+ \varepsilon_{ab}\, 
\epsilon_i\widehat L_i^a\widehat H_u^b 
\label{WRPV}
\end{eqnarray}
where $i,j=1,2,3$ are generation indices, $a,b=1,2$ are $SU(2)$
indices.

\noindent
The set of soft supersymmetry
breaking terms are
\beq
V^{soft}=V^{soft}_{ MSSM}+V^{soft}_{\Slash{R}}
\eeq
\begin{eqnarray}
V^{soft}_{\hbox{\tiny MSSM}}&\hskip -3mm=\hskip -3mm&
M_Q^{ij2}\widetilde Q^{a*}_i\widetilde Q^a_j+M_U^{ij2}
\widetilde U^*_i\widetilde U_j+M_D^{ij2}\widetilde D^*_i 
\widetilde D_j\cr
&\hskip -3mm\hskip -3mm&
\vb{18}
+M_L^{ij2}\widetilde L^{a*}_i\widetilde L^a_j
+M_R^{ij2}\widetilde R^*_i\widetilde R_j \cr
&\hskip -3mm\hskip -3mm&
\vb{18}
+m_{H_d}^2 H^{a*}_d H^a_d+m_{H_u}^2 H^{a*}_u H^a_u \cr
&\hskip -3mm\hskip -3mm&
\vb{20}
- \left[\half \sum_{i=1}^{3} M_i\lambda_i\lambda_i+h.c.\right]\cr
&\hskip -3mm\hskip -3mm&
\vb{20}
+\varepsilon_{ab}\left[
A_U^{ij}\widetilde Q^a_i\widetilde U_j H_u^b
+A_D^{ij}\widetilde Q^b_i\widetilde D_j H_d^a \right.\cr
&\hskip -3mm\hskip -3mm&
\vb{18}
\left. 
+A_E^{ij}\widetilde L^b_i\widetilde R_j H_d^a 
-B\mu H_d^a H_u^b + h.c.\right]
\, ,
\label{SoftMSSM}
\end{eqnarray}
and
\begin{eqnarray}
V^{soft}_{\Slash{R}}&\hskip -3mm=\hskip -3mm&
\varepsilon_{ab}\left[
A_{\lambda}^{ij}\widetilde L^a_i\widetilde L_j^b \widetilde R_k
+A_{\lambda'}^{ijk}\widetilde D_i\widetilde L_j^a \widetilde Q_k^b 
\right] \cr
&\hskip -3mm\hskip -3mm&
\vb{18}
+A_{\lambda^{''}}^{ij}\widetilde D_i\widetilde D_j \widetilde U_k 
+\varepsilon_{ab}\,
B_i\epsilon_i\widetilde L^a_i H_u^b + h.c.
\label{SoftRPV}
\end{eqnarray}

The bilinear
$R_P$ violating term {\sl cannot} be eliminated by superfield
redefinition as sometimes it is claimed. To show this 
we consider the case \cite{marco} where 
all the trilinear couplings in \Eq{WRPV} are zero and 
for simplicity we take $\epsilon_1=\epsilon_2=0$. Then the
superpotential is
\begin{eqnarray} 
W
&\hskip -3mm=\hskip -3mm&
\varepsilon_{ab}\left[
 h_t\widehat Q_3^a\widehat U_3\widehat H_u^b
+h_b\widehat Q_3^b\widehat D_3\widehat H_d^a
+h_{\tau}\widehat L_3^b\widehat R_3\widehat H_d^a \right.\cr
\vb{18}
&\hskip -3mm\hskip -3mm&
\left.
\hskip 1cm
-\mu\widehat H_d^a\widehat H_u^b
+\epsilon_3\widehat L_3^a\widehat H_u^b\right]
\end{eqnarray}
Consider now the rotation
\beq
\widehat H_d'={{\mu\widehat H_d-\epsilon_3\widehat L_3}\over{
\sqrt{\mu^2+\epsilon_3^2}}}\,,\qquad
\widehat L_3'={{\epsilon_3\widehat H_d+\mu\widehat L_3}\over{
\sqrt{\mu^2+\epsilon_3^2}}}
\eeq
In the new basis
\begin{eqnarray} 
W&\hskip -3mm=\hskip -3mm&
h_t\widehat Q_3\widehat U_3\widehat H_u
+h_b{{\mu}\over{\mu'}}\widehat Q_3\widehat D_3\widehat H'_d
+h_{\tau}\widehat L'_3\widehat R_3\widehat H'_d \cr
\vb{18}
&\hskip -3mm\hskip -3mm&
-\mu'\widehat H'_d\widehat H_u 
+h_b{{\epsilon_3}\over{\mu'}}\widehat Q_3\widehat D_3\widehat L'_3
\end{eqnarray}
where
\beq
\mu'^2=\mu^2+\epsilon_3^2
\eeq
But the soft terms,
\begin{eqnarray}
V_{soft}&=&
m_{H_d}^2|H_d|^2+M_{L_3}^2|\widetilde L_3|^2 \cr
&&
\vb{18}
+\left[B\mu H_dH_u \hskip -2pt -\hskip -2pt 
B_2\epsilon_3\widetilde L_3H_u+h.c.\right]\cr
&&
\vb{18}
+ \cdots
\end{eqnarray}
in the rotated basis is
\begin{eqnarray}
V_{soft}&\hskip -11pt=\hskip -11pt&
{{m_{H_d}^2\mu^2\!\!+\!\!M_{L_3}^2\epsilon_3^2}\over{\mu'^2}}|H'_d|^2
\!\!+\!{{m_{H_d}^2\epsilon_3^2\!\!+\!\!M_{L_3}^2\mu^2}\over{\mu'^2}}|\widetilde
L'_3|^2 \cr
\vb{20}
&\hskip -30pt&
\hskip -12pt -\! \left[{{B\mu^2\!+\!B_2\epsilon_3^2}\over{\mu'}}H'_dH_u
\!-\!{{\epsilon_3\mu}\over{\mu'^2}}(m_{H_d}^2\!-\!M_{L_3}^2)\widetilde
L'_3H'_d \right. \cr
\vb{20}
&\hskip -30pt&\left.
\hskip -12pt -{{\epsilon_3\mu}\over{\mu'}}(B_2-B)\widetilde L'_3H_u+h.c.
\right]+ \cdots
\end{eqnarray}
The last two terms violate $R_P$ and 
induce a non-zero VEV for the $\tau$ 
sneutrino field in the rotated basis 
$\left\langle\tilde\nu'_{\tau}\right\rangle=v'_3/\sqrt{2}$, where
\beq
v'_3\approx -{{\epsilon_3\mu}\over{\mu'^2m_{\tilde\nu^0_{\tau}}^2}}
\left(v'_1\Delta m^2+\mu'v_2\Delta B\right)
\eeq
and
\begin{eqnarray}
\Delta m^2&\hskip -3mm\equiv\hskip -3mm& 
m_{H_1}^2 -M_{L_3}^2 \cr
\vb{20}
&\hskip -3mm\approx\hskip -3mm& 
- {{3h_b^2}\over{8\pi^2}}
\left(\vb{14} m_{H_1}^2 \!+\! M_Q^2
\!+\! M_D^2 \!+\!
A_D^2\right)\ln{{M_{GUT}}\over{m_Z}}
\cr
\vb{24}
\Delta B&\hskip -3mm\equiv\hskip -3mm& 
B_2-B\approx{{3h_b^2}\over{8\pi^2}}A_D\ln{{M_{GUT}}\over{m_Z}}
\end{eqnarray}

\section{Bilinear R-Parity Violation}

\subsection{The Model}

The superpotential $W$ for the bilinear $R_P$ violation model is 
given by \cite{e3others,chaHiggsEps}
\begin{eqnarray}
W&\hskip -3mm=\hskip -3mm&
\varepsilon_{ab}\! \left[
 h_U^{ij}\widehat Q_i^a\widehat U_j\widehat H_u^b
\!+\! h_D^{ij}\widehat Q_i^b\widehat D_j\widehat H_d^a
\!+\! h_E^{ij}\widehat L_i^b\widehat R_j\widehat H_d^a \right.\cr
\vb{18}
&\hskip -6mm&
\left. \hskip 1cm 
-\mu\widehat H_d^a\widehat H_u^b
+\epsilon_i\widehat L_i^a\widehat H_u^b\right]
\end{eqnarray}
where $i,j=1,2,3$ are generation indices, $a,b=1,2$ are $SU(2)$
indices.
The set of soft supersymmetry
breaking terms are
\begin{eqnarray}
V_{soft}&\hskip -3mm=\hskip -3mm&
M_Q^{ij2}\widetilde Q^{a*}_i\widetilde Q^a_j+M_U^{ij2}
\widetilde U^*_i\widetilde U_j+M_D^{ij2}\widetilde D^*_i
\widetilde D_j \cr
\vb{18}
&\hskip -6mm&
+M_L^{ij2}\widetilde L^{a*}_i\widetilde L^a_j
+M_R^{ij2}\widetilde R^*_i\widetilde R_j+m_{H_d}^2 H^{a*}_d H^a_d\cr
\vb{18}
&\hskip -6mm&
+m_{H_u}^2 H^{a*}_u H^a_u 
- \left[\half \sum M_i\lambda_i\lambda_i+h.c.\right]\cr
\vb{18}
&\hskip -6mm&
+\varepsilon_{ab}\left[
A_U^{ij}\widetilde Q^a_i\widetilde U_j H_u^b
+A_D^{ij}\widetilde Q^b_i\widetilde D_j H_d^a\right.\cr
\vb{18}
&\hskip -6mm& \left.
+A_E^{ij}\widetilde L^b_i\widetilde R_j H_d^a 
\!-\!B\mu H_d^a H_u^b+B_i\epsilon_i\widetilde L^a_i H_u^b\right]
\,.\cr
\vb{12}
&\hskip -6mm&
\end{eqnarray}

\noindent
The electroweak symmetry is broken when the VEVS of 
the two Higgs doublets $H_d$
and $H_u$, and the sneutrinos.
\begin{eqnarray}
H_d&=&{{{1\over{\sqrt{2}}}[\chi^0_d+v_d+i\varphi^0_d]}\choose{
H^-_d}} \\
H_u&=&{{H^+_u}\choose{{1\over{\sqrt{2}}}[\chi^0_u+v_u+
i\varphi^0_u]}}\\
L_i&=&{{{1\over{\sqrt{2}}}
[\tilde\nu^R_{i}+v_i+i\tilde\nu^I_{i}]}\choose{\tilde\ell^{i}}}
\end{eqnarray}
The gauge bosons $W$ and $Z$ acquire masses
\beq
m_W^2=\quarter g^2v^2 \quad ; \quad m_Z^2=\quarter(g^2+g'^2)v^2
\eeq
where
\beq
v^2\equiv v_d^2+v_u^2+v_1^2+v_2^2+v_3^2=(246 \; {\rm GeV})^2
\eeq
We introduce the
following notation in spherical coordinates:
\begin{eqnarray}
v_d&=&v\sin\theta_1\sin\theta_2\sin\theta_3\cos\beta\cr
v_u&=&v\sin\theta_1\sin\theta_2\sin\theta_3\sin\beta\cr
v_1&=&v\sin\theta_1\sin\theta_2\cos\theta_3\cr
v_2&=&v\sin\theta_1\cos\theta_2\cr
v_3&=&v\cos\theta_1\nn
\end{eqnarray}
which preserves the MSSM 
definition $\tan\beta=v_u/v_d$. The angles $\theta_i$
are equal to $\pi/2$ in the MSSM limit.

\noindent
The full scalar potential may be written as

\beq
V_{total}  = \sum_i \left| { \partial W \over \partial z_i} \right|^2
+ V_D + V_{soft} + V_{RC}
\eeq
where $z_i$ denotes any one of the scalar fields in the
theory, $V_D$ are the usual $D$-terms, $V_{soft}$ the SUSY soft
breaking terms, and $V_{RC}$ are the 
one-loop radiative corrections. 

\noindent
In writing $V_{RC}$ we  use the diagrammatic method and find 
the minimization conditions by correcting to one--loop the tadpole
equations. 
This method has advantages with respect to the effective potential when
we calculate the one--loop corrected scalar masses.
The scalar potential contains linear terms
\beq
V_{linear}=t_d\sigma^0_d+t_u\sigma^0_u+t_i\tilde\nu^R_{i}
\equiv t_{\alpha}\sigma^0_{\alpha}\,,
\eeq
where we have introduced the notation
\beq
\sigma^0_{\alpha}=(\sigma^0_d,\sigma^0_u,\nu^R_1,\nu^R_2,\nu^R_3)
\eeq
and $\alpha=d,u,1,2,3$. The one loop tadpoles are
\begin{eqnarray}
t_{\alpha}&=&t^0_{\alpha} -\delta t^{\overline{MS}}_{\alpha}
+T_{\alpha}(Q)\cr
\vb{22}
&=&t^0_{\alpha} +T^{\overline{MS}} _{\alpha}(Q)
\label{tadpoles}
\end{eqnarray}
where $T^{\overline{MS}} _{\alpha}(Q)\equiv -\delta t^{\overline{MS}}_{\alpha}
+T_{\alpha}(Q)$ are the finite one--loop tadpoles.

\subsection{Main Features}

The $\epsilon$--model is a one (three) parameter(s) generalization of
the MSSM.
It can be thought as an effective model
showing the more important features of the SBRP--model~\cite{SBRpV} 
at the weak
scale. 
The mass matrices, charged and neutral currents, are similar to the
SBRP--model if we identify
\beq
\epsilon \equiv v_R h_{\nu}
\eeq
The $R_P$ violating
parameters $\epsilon_i$ and $v_i$ violate lepton number, inducing
a non-zero mass for only one neutrino, which could be considered
to be the the $\nu_{\tau}$. 
The $\nu_e$ and $\nu_{\mu}$
remain massless in first approximation.  As we will explain below, 
they acquire 
masses from supersymmetric loops \cite{numass,ralf} that are typically
smaller than the tree level mass.

\begin{figure}
\begin{picture}(0,8)
\put(0,0){\includegraphics[width=7cm]{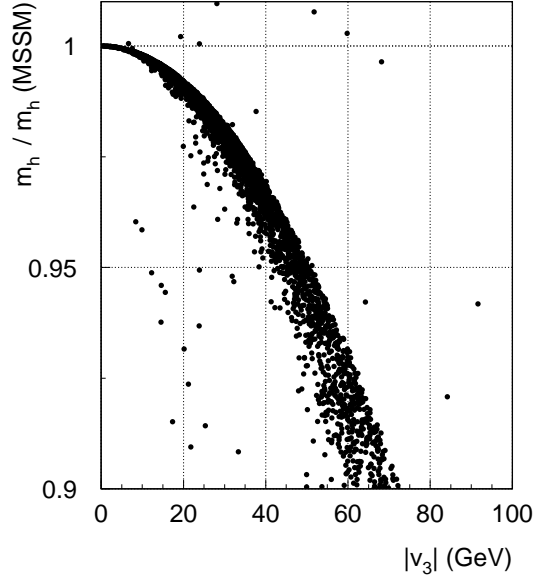}}
\end{picture}
\vspace{-10mm}
\caption{Ratio of the lightest CP-even Higgs
boson mass $m_h$ in the
$\epsilon$--model and in the MSSM  as a function of
$v_3$.}
\label{fig1}
\end{figure}

The model has the MSSM as a limit. This can be illustrated in
Figure~\ref{fig1} where we show the ratio of the lightest CP-even Higgs
boson mass $m_h$ in the
$\epsilon$--model and in the MSSM  as a function of
$v_3$. 
Many other results concerning this model and the implications for
physics at the accelerators can be found in ref.~\cite{e3others,chaHiggsEps}.

\section{Radiative Breaking}

\subsection{Radiative Breaking in the $\epsilon$ model: The minimal case}

At $Q = M_{GUT}$ we assume the standard minimal supergravity
unifications assumptions, 
\beqa
&&A_t = A_b = A_{\tau} \equiv A \:, \cr
&&\vb{18}
B=B_2=A-1 \:, \cr
&&\vb{18}
m_{H_d}^2 = m_{H_u}^2 = M_{L}^2 = M_{R}^2 = m_0^2 \:, \cr
&&\vb{18}
M_{Q}^2 =M_{U}^2 = M_{D}^2 = m_0^2 \:, \cr
&&\vb{18}
M_3 = M_2 = M_1 = M_{1/2} \nn
\eeqa
In order to determine the values of the Yukawa couplings and of the
soft breaking scalar masses at low energies we first run the RGE's from
the unification scale $M_{GUT} \sim 10^{16}$ GeV down to the weak
scale. We randomly give values at the unification scale
for the parameters of the theory. 
\beq
\begin{array}{ccccc}
10^{-2} & \leq &{h^2_t}_{GUT} / 4\pi & \leq&1 \cr
10^{-5} & \leq &{h^2_b}_{GUT} / 4\pi & \leq&1 \cr
-3&\leq&A/m_0&\leq&3 \cr
0&\leq&\mu^2_{GUT}/m_0^2&\leq&10 \cr
0&\leq&M_{1/2}/m_0&\leq&5 \cr
10^{-2} &\leq& {\epsilon^2_i}_{GUT}/m_0^2 &\leq& 10\cr 
\end{array}
\eeq
The values of $h_{e}^{GUT},h_{\mu}^{GUT},h_{\tau}^{GUT}$ are 
defined in such a way
that we get the charged lepton  masses correctly. 
As the charginos mix with the 
leptons, through a mass matrix given by
\beq
{\cal M}_C=\left[
\matrix{ M_C & A \cr
\vb{18}
B & M_L}
\right]
\eeq
where $M_C$ is the usual MSSM chargino mass matrix,
\beq
M_C=\left[
\matrix{ M & {\textstyle{1\over{\sqrt{2}}}}gv_u\cr
\vb{18}
{\textstyle{1\over{\sqrt{2}}}}gv_d & \mu}
\right]
\eeq
$M_L$ is the lepton mass matrix, that we consider diagonal,
\beq
M_L=\left[
\matrix{ {\textstyle{1\over{\sqrt{2}}}}h_{E_{11}}v_d & 0 & 0 \cr
\vb{18}
0& {\textstyle{1\over{\sqrt{2}}}}h_{E_{22}}v_d & 0 \cr
\vb{18}
0 &0 &{\textstyle{1\over{\sqrt{2}}}}h_{E_{33}}v_d}
\right]
\eeq
and $A$ and $B$ are matrices that are non zero due to the violation of
$R_P$ and are given by
\beq
A^T=\hskip -3pt\left[
\matrix{ -\half h_{E_{11}}v_1 & 0\cr
\vb{18}
-\half h_{E_{22}}v_1 & 0\cr
\vb{18}
-\half h_{E_{33}}v_3 & 0}
\right]
B=\hskip -3pt\left[
\matrix{ \half gv_3 & -\epsilon_1\cr
\vb{18}
\half gv_3 & -\epsilon_2\cr
\vb{18}
\half gv_3 & -\epsilon_3}
\right]
\eeq
We used an iterative procedure to accomplish that the three lightest
eigenvalues of ${\cal M}_C$ are in agreement with  
the experimental masses of the leptons.
After running the RGE we have a 
complete set of parameters, Yukawa couplings and soft-breaking masses 
$m^2_i(RGE)$ to study the minimization. 
This is done by the following method: we solve the minimization
equations for the soft masses squared. This is easy because those
equations are linear on the soft masses squared. The values obtained
in this way, that we call $m^2_i$ are not equal to the values
$m^2_i(RGE)$ that we got via RGE. To achieve equality we define a function
\beq
\eta= max \left(  \frac{\ds m^2_i}{\ds m^2_i(RGE)},
\frac{\ds m^2_i(RGE)}{\ds m^2_i}
\right) \quad \forall i 
\eeq
with the obvious property that
\beq
\eta \ge 1
\eeq
Then we adjust the parameters to minimize $\eta$.

Before we end this section let us discuss the counting of free
parameters in this model and in the minimal N=1 supergravity unified
version of the MSSM. In \Table{table1} we show this counting for the
MSSM and in \Table{table2} for the $\epsilon$--model.
Finally, we note that in either case, the sign of the mixing parameter
$\mu$ is physical and has to be taken into account.

\begin{table}
\caption{Counting of free parameters in MSSM}
\begin{tabular}{ccc}\hline
Parameters  
\hskip -9pt&\hskip -9pt 
Conditions 
\hskip -9pt&\hskip -9pt 
Free Parameters \cr \hline
$h_t$, $h_b$, $h_{\tau}$
\hskip -9pt&\hskip -9pt 
$m_W$, $m_t$
\hskip -9pt&\hskip -9pt 
 $\tan\beta$ \cr 
$v_d$, $v_u$,$M_{1/2}$ 
\hskip -9pt&\hskip -9pt 
$m_b$, $m_{\tau}$ 
\hskip -9pt&\hskip -9pt 
 2 Extra \cr 
 $m_0$, $A$, $\mu$
\hskip -9pt&\hskip -9pt 
$t_i=0$, $i=1,2$
\hskip -9pt&\hskip -9pt 
({\it e.g.} $m_h$, $m_A$)\cr \hline
Total = 9
\hskip -9pt&\hskip -9pt 
Total = 6 
\hskip -9pt&\hskip -9pt 
Total = 3\cr\hline
\end{tabular}
\label{table1}
\end{table}

\begin{table}
\caption{Counting of free parameters in our model}
\begin{tabular}{ccc}\hline
Parameters 
\hskip -9pt&\hskip -9pt 
 Conditions 
\hskip -9pt&\hskip -9pt 
 Free Parameters \cr \hline
$h_t$, $h_b$, $h_{\tau}$
\hskip -9pt&\hskip -9pt 
$m_W$, $m_t$ 
\hskip -9pt&\hskip -9pt 
 $\tan\beta$, $\epsilon_i$ \cr 
$v_d$, $v_u$, $M_{1/2}$
\hskip -9pt&\hskip -9pt 
$m_b$, $m_{\tau}$ 
\hskip -9pt&\hskip -9pt  \cr 
$m_0$,$A$, $\mu$
\hskip -9pt&\hskip -9pt 
$t_i=0$
\hskip -9pt&\hskip -9pt 
 2 Extra \cr 
$v_i$, $\epsilon_i$
\hskip -9pt&\hskip -9pt 
($i=1,\ldots,5$)
\hskip -9pt&\hskip -9pt 
 ({\it e.g.} $m_h$, $m_A$)\cr \hline
Total = 15
\hskip -9pt&\hskip -9pt 
Total = 9 
\hskip -9pt&\hskip -9pt 
Total = 6\cr\hline
\end{tabular}
\label{table2}
\end{table}

\subsection{Gauge and Yukawa Unification in the $\epsilon$ model}

There is a strong motivation to consider GUT theories where {\it
both} gauge and Yukawa unification can achieved. This is because 
besides achieving gauge coupling unification,
GUT theories also reduce the number of free parameters in the Yukawa
sector and this is normally a desirable feature. The situation with
respect to the MSSM can be summarized as follows:

\vspace{-1mm}

\begin{itemize}
\item
In $SU(5)$ models, $h_b=h_{\tau}$ at $M_{GUT}$.
The predicted ratio $m_b/m_{\tau}$ at $M_{WEAK}$ agrees with
experiments. 

\vspace{-1mm}

\item
A relation between $m_{top}$ and
$\tan\beta$ is predicted. Two solutions are possible: low and high 
$\tan\beta$ .

\vspace{-1mm}

\item
In $SO(10)$ and $E_6$ models $h_t=h_b=h_{\tau}$ at $M_{GUT}$.
In this case, only the large
$\tan\beta$ solution survives.

\end{itemize}

\vspace{-1mm}

\noindent
We have shown~\cite{YukUnifBRpV} that the $\epsilon$--model allows $b-\tau$ Yukawa unification for
any value of $\tan\beta$ and satisfying perturbativity of the
couplings.  We also find the $t-b-\tau$ Yukawa unification 
easier to achieve than in the MSSM, occurring in a 
wider high $\tan\beta$ region.

\section{Tree Level Neutrino Masses and Mixings}

\subsection{Neutral fermion mass matrix}

In the basis 
\beq
\psi^{0T}= 
(-i\lambda',-i\lambda^3,\widetilde{H}_d^1,\widetilde{H}_u^2,
\nu_{e}, \nu_{\mu}, \nu_{\tau} )
\eeq
the neutral fermions mass terms in the Lagrangian are given by 
\beq
{\cal L}_m=-\frac 12(\psi^0)^T{\bold M}_N\psi^0+h.c.   
\eeq
where the neutralino/neutrino mass matrix is 
\beq
{\bold M}_N=\left[  
\begin{array}{cc}  
{\cal M}_{\chi^0}& m^T \cr
\vb{12}
m & 0 \cr
\end{array}
\right]
\eeq
with
\beq
{\cal M}_{\chi^0}\hskip -2pt=\hskip -4pt \left[ \hskip -7pt 
\begin{array}{cccc}  
M_1 & 0 & -\frac 12g^{\prime }v_d & \frac 12g^{\prime }v_u \cr
\vb{12}   
0 & M_2 & \frac 12gv_d & -\frac 12gv_u \cr
\vb{12}   
-\frac 12g^{\prime }v_d & \frac 12gv_d & 0 & -\mu  \cr
\vb{12}
\frac 12g^{\prime }v_u & -\frac 12gv_u & -\mu & 0  \cr
\end{array}  
\hskip -6pt
\right] 
\eeq
\beq
m=\left[  
\begin{array}{cccc}  
-\frac 12g^{\prime }v_1 & \frac 12gv_1 & 0 & \epsilon _1 \cr
\vb{12}
-\frac 12g^{\prime }v_2 & \frac 12gv_2 & 0 & \epsilon _2  \cr
\vb{12}
-\frac 12g^{\prime }v_3 & \frac 12gv_3 & 0 & \epsilon _3  \cr  
\end{array}  
\right] 
\eeq

\noindent 
The mass matrix ${\bold M}_N$ is diagonalized by 
\beq
{\cal  N}^*{\bold M}_N{\cal N}^{-1}={\rm diag}(m_{\chi^0_i},m_{\nu_j})
\label{eq:NeuMdiag} 
\eeq
where $(i=1,\cdots,4)$ for the neutralinos, and $(j=1,\cdots,3)$ for
the neutrinos.

\subsection{Approximate diagonalization of mass matrices}

If the \rp parameters are small it is convenient to define~\cite{MartinValle} the matrix
\beq
\xi = m \cdot {\cal M}_{\chi^0}^{-1}
\eeq
If the elements of this matrix satisfy
\beq
\forall \xi_{ij} \ll 1
\eeq
then one can find an approximate solution for 
mixing  matrix ${\cal N}$.
Explicitly we have
\begin{eqnarray}
\xi_{i1} &=& \frac{g' M_2 \mu}{2 det({\cal M}_{\chi^0})}\Lambda_i \cr
\vb{20}
\xi_{i2} &=& -\frac{g M_1 \mu}{2 det({\cal M}_{\chi^0})}\Lambda_i \cr
\vb{20}
\xi_{i3} &=& - \frac{\epsilon_i}{\mu} + 
          \frac{(g^2 M_1 + {g'}^2 M_2) v_2}
               {4 det({\cal M}_{\chi^0})}\Lambda_i \cr
\vb{20}
\xi_{i4} &=& - \frac{(g^2 M_1 + {g'}^2 M_2) v_1}
               {4 det({\cal M}_{\chi^0})}\Lambda_i
\label{xielementos}
\end{eqnarray}
where
\beq
\Lambda_i = \mu v_i + v_d \epsilon_i
\label{lambdai}
\eeq

\noindent 
From \Eq{xielementos} and \Eq{lambdai} one can see that $\xi=0$ in the
MSSM limit where $\epsilon_i=0$, $v_i=0$.
In leading order in $\xi$ the mixing matrix ${\cal N}$ is given by,
\beq
{\cal N}^* \hskip -2pt =\hskip -2pt  \left(\hskip -1pt
\begin{array}{cc}
N^* & 0\\
0& V_\nu^T \end{array}
\hskip -1pt
\right)
\left(
\hskip -1pt
\begin{array}{cc}
1 -{1 \over 2} \xi^{\dagger} \xi& \xi^{\dagger} \\
-\xi &  1 -{1 \over 2} \xi \xi^\dagger
\end{array}
\hskip -1pt
\right) 
\eeq

\noindent
The second matrix above block-diagonalizes 
${\bold M}_N$ approximately to the form 
diag($m_{eff},{\cal M}_{\chi^0}$)

\beqa
m_{eff} &\hskip -3mm=\hskip -3mm& 
- m \cdot {\cal M}_{\chi^0}^{-1} m^T \cr
\vb{18}
&\hskip -3mm=\hskip -3mm& 
\frac{M_1 g^2 \!+\! M_2 {g'}^2}{4\, det({\cal M}_{\chi^0})} 
\left(\hskip -2mm \begin{array}{ccc}
\Lambda_e^2 
\hskip -1pt&\hskip -1pt
\Lambda_e \Lambda_\mu
\hskip -1pt&\hskip -1pt
\Lambda_e \Lambda_\tau \\
\Lambda_e \Lambda_\mu 
\hskip -1pt&\hskip -1pt
\Lambda_\mu^2
\hskip -1pt&\hskip -1pt
\Lambda_\mu \Lambda_\tau \\
\Lambda_e \Lambda_\tau 
\hskip -1pt&\hskip -1pt 
\Lambda_\mu \Lambda_\tau 
\hskip -1pt&\hskip -1pt
\Lambda_\tau^2
\end{array}\hskip -3mm \right)
\eeqa

\noindent
The sub-matrices $N$ and $V_{\nu}$ in eq.  diagonalize 
${\cal M}_{\chi^0}$ and $m_{eff}$ 
\beq
N^{*}{\cal M}_{\chi^0} N^{\dagger} = {\rm diag}(m_{\chi^0_i}),
\eeq
\beq
V_{\nu}^T m_{eff} V_{\nu} = {\rm diag}(0,0,m_{\nu}),
\eeq
where 
\beq
m_{\nu} = Tr(m_{eff}) = 
\frac{M_1 g^2 + M_2 {g'}^2}{4\, det({\cal M}_{\chi^0})} 
|{\vec \Lambda}|^2.
\eeq

\noindent
For $V_{\nu}$ we have ( we can rotate away one angle)

\beqa
V_{\nu}&=& 
\left(\begin{array}{ccc}
  1 &                0 &               0 \\
  0 &  \cos\theta_{23} & -\sin\theta_{23} \\
  0 &  \sin\theta_{23} & \cos\theta_{23} 
\end{array}\right) \times \cr
\vb{28}
&&
\left(\begin{array}{ccc}
  \cos\theta_{13} & 0 & -\sin\theta_{13} \\
                0 & 1 &               0 \\
  \sin\theta_{13} & 0 & \cos\theta_{13} 
\end{array}\right) ,
\eeqa
where the mixing angles can be expressed in terms of the 
{\it alignment vector} ${\vec \Lambda}$ as follows:
\beq
\tan\theta_{13} = - \frac{\Lambda_e}
                   {(\Lambda_{\mu}^2+\Lambda_{\tau}^2)^{\frac{1}{2}}},
\eeq
\beq
\tan\theta_{23} = \frac{\Lambda_{\mu}}{\Lambda_{\tau}}.
\eeq

\section{One Loop Neutrino Masses and Mixings}

\subsection{Definition}

The Self--Energy for the neutralino/neutrino is

\vbox{
\beqa
\hskip 2.7cm 
&\equiv& \hskip -1mm
i \left\{ \vb{14}\slash{p} \left[\vb{12} 
P_L \Sigma^L_{ij} + P_R \Sigma^R_{ij}
\right] \right.\cr
\vb{20}
&&\left.\hskip -1mm \vb{14}
-\left[ \vb{12}
P_L \Pi^L_{ij} + P_R \Pi^R_{ij} \right]\right\}
\eeqa

\begin{picture}(0,0)
\put(-0.3,0.95){\includegraphics[width=3cm]{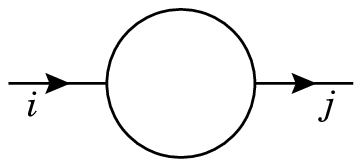}}
\end{picture}
}

\noindent
Then 
\beq
M^{\rm pole}_{ij}= M^{\rm \overline{DR}}_{ij}(\mu_R) + \Delta M_{ij}
\eeq
where
\beqa
\Delta M_{ij}&\hskip -3mm=\hskip -3mm& \left[ \half 
\left(\Pi^V_{ij}(m_i^2) + \Pi^V_{ij}(m_j^2)\right) \right.\cr
\vb{18}
&\hskip -6mm&\left.
\!- \half 
\left( m_{\chi^0_i} \Sigma^V_{ij}(m_i^2)\!  +\!
m_{\chi^0_j} \Sigma^V_{ij}(m_j^2) \right) \right]_{\Delta=0}
\eeqa
where
\beq
\Sigma^V=\half \left(\Sigma^L+\Sigma^R\right)
\qquad
\Pi^V=\half \left(\Pi^L+\Pi^R\right)
\eeq
and
\beq
\ds \Delta=\frac{2}{4-d} -\gamma_E + \ln 4\pi
\eeq

\subsection{Diagrams Contributing}

\begin{figure}
\begin{picture}(0,4.5)
\put(0,2.5){\includegraphics{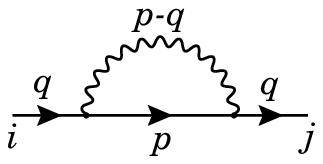}}
\end{picture}
\begin{picture}(0,4.5)
\put(3.5,2.5){\includegraphics{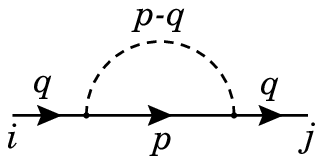}}
\end{picture}
\begin{picture}(0,3)
\put(0,-0.45){\includegraphics{sself.eps}}
\end{picture}
\begin{picture}(0,3)
\put(3.5,0){\includegraphics{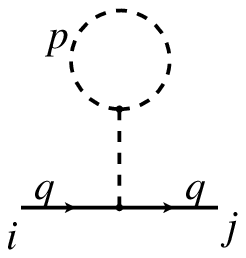}}
\end{picture}
\vspace{-5mm}
\caption{Diagrams contributing}
\label{fig2}
\end{figure}

In \Fig{fig2} are shown the classes of diagrams contributing to the
self--energy at one loop.
These diagrams can be calculated in a straightforward way. For
instance the $W$ diagram in the $\xi=1$ gauge gives

\beqa
\Sigma^V_{ij}
&\hskip -5pt=\hskip -5pt& 
-\frac{1}{16\pi^2}\, \sum_{k=1}^5
2 \left(O^{\rm ncw}_{L jk} O^{\rm cnw}_{L ki} +
O^{\rm ncw}_{R jk} O^{\rm cnw}_{R ki}\right) \cr
\vb{14}
&\hskip -15pt \hskip -15pt& 
\hskip 2cm B_1(p^2,m^2_k,m^2_W)\cr
\vb{28}
\Pi^V_{ij}
&\hskip -5pt=\hskip -5pt& 
 -\frac{1}{16\pi^2}\, \sum_{k=1}^5
(-4) \left(O^{\rm ncw}_{L jk} O^{\rm cnw}_{R ki} +
O^{\rm ncw}_{R jk} O^{\rm cnw}_{L ki}\right)\cr 
&\hskip -15pt \hskip -15pt& 
\vb{14}
\hskip 2cm m_k\, B_0(p^2,m^2_k,m^2_W)
\eeqa
where $B_0$ and $B_1$ are the Passarino-Veltman functions, and
the coupling matrices appear in the vertices in the following way

\vbox{
\vspace{8mm}
\beq
\hskip 3cm i\ \gamma^{\mu} \left( O^{\rm ncw}_{L ji} P_L +
O^{\rm ncw}_{R ji} P_R \right)
\eeq

\begin{picture}(0,1.5)
\put(0,0.25){\includegraphics[height=3cm]{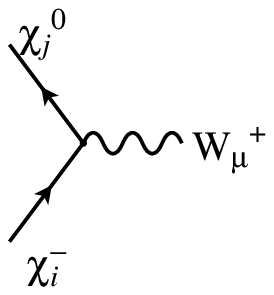}}
\end{picture}
}

\beq
\hskip 3cm i\ \gamma^{\mu} \left( O^{\rm cnw}_{L ji} P_L +
O^{\rm cnw}_{R ji} P_R \right)
\eeq

\begin{picture}(0,0)
\put(0,-1){\includegraphics[height=3cm]{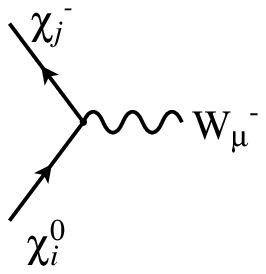}}
\end{picture}

\vspace{5mm}

\subsection{Gauge Invariance}

\begin{figure}
\begin{picture}(0,2.5)
\put(0,0){\includegraphics[width=3cm]{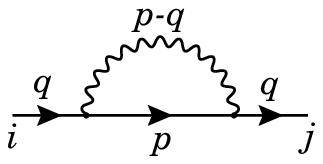}}
\end{picture}
\begin{picture}(0,0)
\put(2.5,0){\includegraphics[width=3cm]{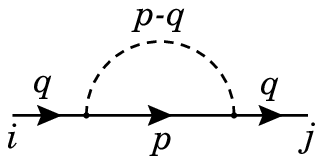}}
\end{picture}
\begin{picture}(0,0)
\put(5,0.2){\includegraphics[width=3cm]{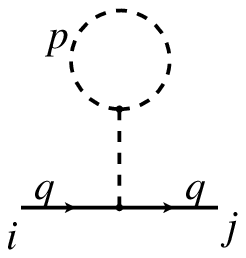}}
\end{picture}
\vspace{-10mm}
\caption{Set 1}
\label{fig3}
\end{figure}
\begin{figure}
\begin{picture}(0,2.5)
\put(1,0){\includegraphics[width=3cm]{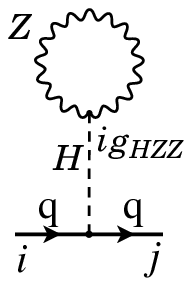}}
\end{picture}
\begin{picture}(0,0)
\put(4,0){\includegraphics[width=3cm]{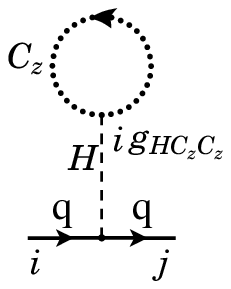}}
\end{picture}
\vspace{-10mm}
\caption{Set 2}
\label{fig4}
\end{figure}
\begin{figure}
\begin{picture}(0,2.5)
\put(0.5,0){\includegraphics[width=3cm]{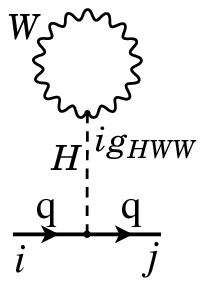}}
\end{picture}
\begin{picture}(0,0)
\put(2.4,0){\includegraphics[width=3cm]{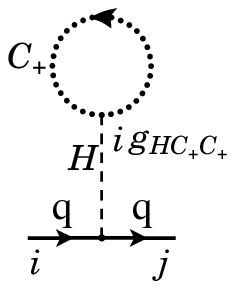}}
\end{picture}
\begin{picture}(0,0)
\put(4.5,0){\includegraphics[width=3cm]{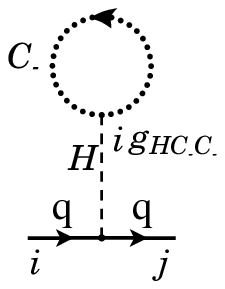}}
\end{picture}
\vspace{-10mm}
\caption{Set 3}
\label{fig5}
\end{figure}

When calculating the self--energies the question of gauge invariance
arises. In the $R_{\xi}$ gauge the sets of diagrams of 
Figs.~\ref{fig3}-\ref{fig5} depend on $\xi$.
We have shown that the gauge dependence cancels among
the diagrams in each set. So in the actual calculations we considered
the tadpoles needed in those sets to ensure gauge invariance. The
other tadpoles were included in the minimization procedure of
\Eq{tadpoles}. This is a gauge invariant splitting.

\subsection{The One--Loop Mass Matrix}

The one--loop corrected mass matrix is given by
\beq
M^{1L}= M^{0L}_{diag} + \Delta M^{1L}
\eeq
where
\beq
M^{0L}_{diag}= {\cal N} M_N {\cal N}^T
\eeq

\noindent
Now we diagonalize the 1--loop mass matrix 
\beq
M^{1L}_{diag}={\cal N'} M^{1L} {\cal N'}^T
\eeq
Then the mass eigenstates are related to the weak basis states by
\beq
\chi_0^{mass}= {\cal N}^{1L}_{i \alpha}\, \chi_0^{weak}
\eeq
with
\beq
{\cal N}^{1L} = {\cal N'}\ {\cal N}
\eeq

\noindent
The usual convention in neutrino physics
\beq
\nu_{\alpha} = U_{\alpha k}\,  \nu_k
\eeq
is recovered in our notation as
\beq
U_{\alpha k}= {\cal N}^{1L}_{4+k, 4+\alpha} 
\eeq

\subsection{Solar and Atmospheric Neutrino Parameters}

\noindent
Assuming hierarchy in the masses $m_{\nu_2}$ and $m_{\nu_3}$ the
survival probabilities for the solar and atmospheric neutrinos 
are

\begin{eqnarray}
P_e&\hskip -3mm=\hskip -3mm&
1 \!-\! 4 U_{e1}^2 U_{e2}^2\ \sin^2 \left( \frac{\Delta m^2_{21}t}{4 E}
\right)
\!-\! 2 U_{e3}^2 (1 - U_{e3}^2 )\cr
\vb{20}
P_{\mu}&\hskip -3mm=\hskip -3mm&
1 \!-\! 4 U_{\mu 3}^2 (1 -U_{\mu 3}^2)\ 
\sin^2 \left( \frac{\Delta m^2_{21}t}{4 E}
\right)
\end{eqnarray}

\noindent
As $U_{e3}$ has to be small we neglect it and write the usual two
neutrino mixing angle as
\beq
\sin^2 (2 \theta_{12}) = 4\, U_{e1}^2 U_{e2}^2
\eeq
and
\beq
\sin^2 (2 \theta_{13}) = 4\, U_{\mu 3}^2 (1 -U_{\mu 3}^2)
\eeq

\subsection{Our Preliminary Results}

We have found~\cite{numass}, that if $\epsilon^2 /\Lambda \ll 100 $
then the approximate formulas hold
\begin{eqnarray}
U_{e3}&\approx&\sin\left
(\tan^{-1} \left(\Lambda_1 / \sqrt{\Lambda_2^2 + \Lambda_3^2}
\right)\right)\cr
\vb{24}
U_{\mu 3}&\approx&\sin\left
(\tan^{-1} \left( \Lambda_2 / \sqrt{\Lambda_1^2 + \Lambda_3^2}
\right)\right)\\
\vb{20}
U_{\tau 3}&\approx&\sin\left
(\tan^{-1} \left( \Lambda_3 / \sqrt{\Lambda_1^2 + \Lambda_2^2}
\right)\right) \nn
\end{eqnarray}

\begin{figure}
\noindent
$\sin^2(2\theta_{atm})$

\begin{picture}(0,4)
\put(0.5,0){\includegraphics[width=70mm]{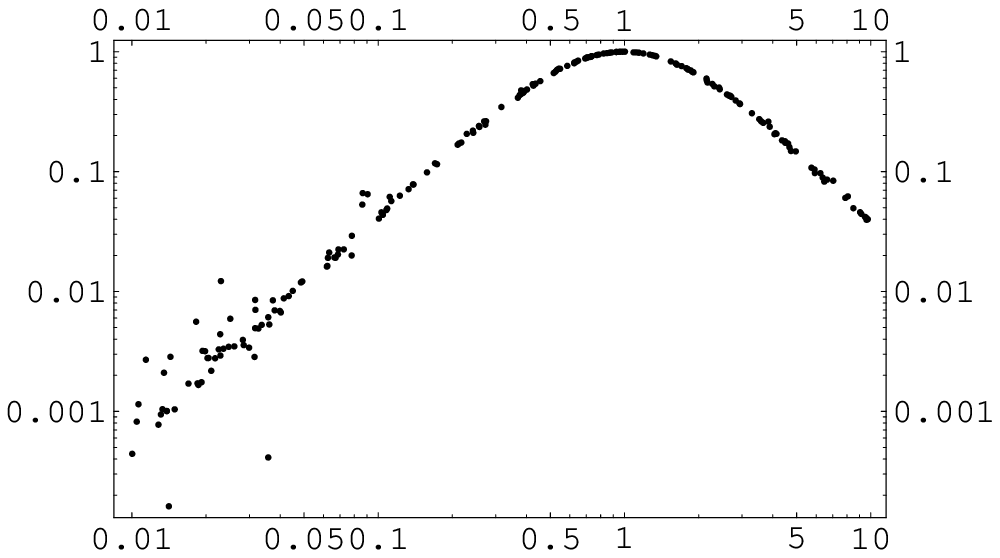}}
\end{picture}
\vspace{1mm}

$\hskip 55mm \Lambda_{\mu}/\Lambda_{\tau}$

\vspace{-6mm} 
\caption{The atmospheric angle as function of
$\Lambda_{\mu}/\Lambda_{\tau}$ for $|\epsilon_i|=\epsilon$ and
$\Lambda_e=0.1\Lambda_{\tau}$. $\epsilon^2/\Lambda$ has an upper cut of
$\epsilon^2/\Lambda \le 0.1$ in this plot, since larger values lead to
larger scatter for very small $\Lambda_{\mu}/\Lambda_{\tau}$}
\label{atm05}
\end{figure}

\noindent
Then if we take
\beq
\Lambda_1 \ll \Lambda_2 \simeq \Lambda_3
\eeq
we immediately get maximal mixing for the atmospheric neutrinos. This
is shown in Figure~\ref{atm05} where we see that maximality of the
mixing is only possible for $\Lambda_{\mu}=\Lambda_{\tau}$.

%
%

\begin{figure}
\noindent
$\sin^2(2\theta_{sol})$

\begin{picture}(0,4)
\put(0.5,0){\includegraphics[width=70mm]{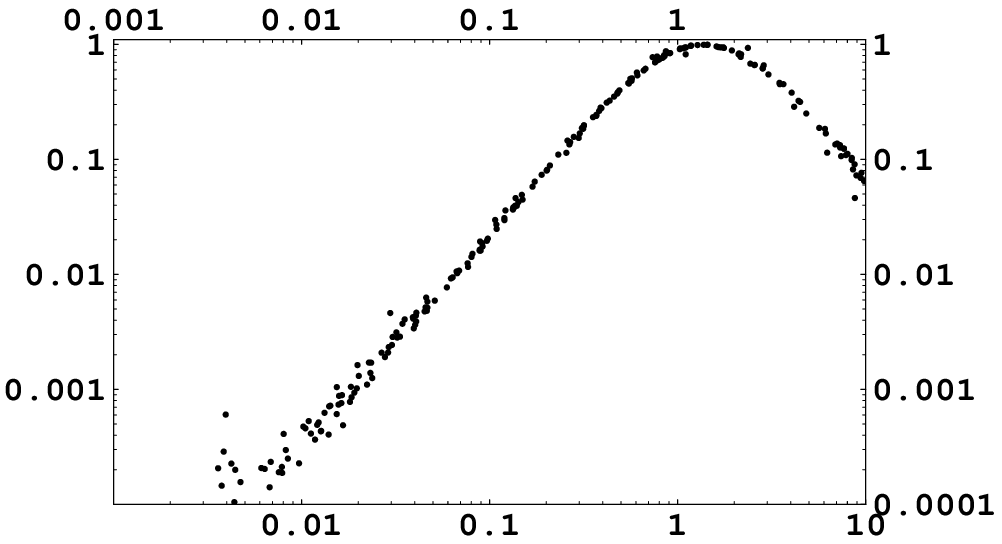}}
\end{picture}
\vspace{1mm}

$\hskip 55mm \epsilon_e/\epsilon_{\mu}$

\vspace{-6mm} 
\caption{The solar angle as function of
$\epsilon_e/\epsilon_{\mu}$ for $\epsilon_{\mu}=\epsilon_{\tau}$ and
$\Lambda_{\mu}=\Lambda_{\tau}$ applying the sign condition.}
\label{solar04}
\end{figure}

To get bimaximality we have to fix the solar angle. We have discovered
that if $\epsilon_e\simeq\epsilon_{\mu}\simeq\epsilon_{\tau}$ and
$\Lambda_e\ll \Lambda_{\mu}\simeq\Lambda_{\tau}$ we get
bimaximality if the following sign condition applies
\beq
(\epsilon_{\mu}/\epsilon_{\tau})\times 
(\Lambda_{\mu}/\Lambda_{\tau}) \le 0
\eeq
This is illustrated in Figure~\ref{solar04}.
In practice we do not need perfect maximality.  We took

%
%

\begin{figure}

\noindent
$\Delta m^2_{23}$

\noindent
$[eV^2]$

\begin{picture}(0,4)
\put(0.5,0){\includegraphics[width=70mm]{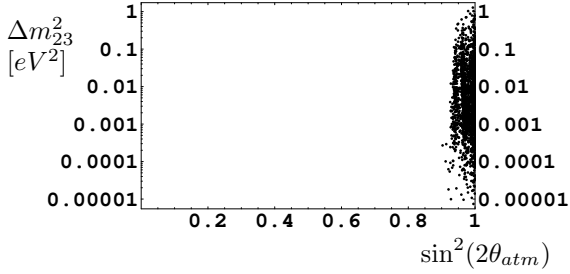}}
\end{picture}
\vspace{-20mm}

$\hskip 55mm \sin^2(2\theta_{atm})$

\vspace{-6mm} 
\caption{$\Delta m^2_{23}$ versus $\sin^2(2\theta_{atm})$.
All points obey Eqs.~(\ref{AtmAngle}), (\ref{SolAngle} and
(\ref{AtmMass}) no further cut applied except that
$0.3 \le \epsilon^2/|\Lambda| \le 1$.}
\label{fig8}
\end{figure}

\beq
4/5 \le |\Lambda_{\mu}/\Lambda_{\tau}| \le 5/4 
\quad \hbox{and} \quad
|\Lambda_{e}/\Lambda_{\mu}| \ll 1
\label{AtmAngle}
\eeq 
to fix the atmospheric angle and
\beq
(\epsilon_{\mu}/\epsilon_{\tau})\times (\Lambda_{\mu}/\Lambda_{\tau}) 
\le 0 \ \hbox{and}\
 0.6 \le \epsilon_{e}/\epsilon_{\mu} \le 1.2
\label{SolAngle}
\eeq
to fix the solar angle.
Next we have to fix the masses to solve the atmospheric and solar
neutrino problems. We found~\cite{numass,MartinValle} that the range
\beq
0.03 {\rm GeV}^2 \le |\Lambda| \le  0.25 {\rm GeV}^2
\label{AtmMass}
\eeq
fixes the tree level mass to reproduce the atmospheric neutrino problem.
This is illustrated in Figure~\ref{fig8} where, besides the conditions in
Eqs.(\ref{AtmAngle}) and (\ref{SolAngle}) to fix the angles and
condition Eq.~(\ref{AtmMass}), all the other parameters were
chosen randomly. Consistency of the parameters was required in the
sense that minimization the scalar potential including the
tadpoles was performed as well as the matching with the RGE solutions 
with universality at GUT scale.

%
%

%
%

\begin{figure}

\noindent
$\Delta m^2_{12}$

\noindent
$[eV^2]$

\begin{picture}(0,4)
\put(0.5,0){\includegraphics[width=70mm]{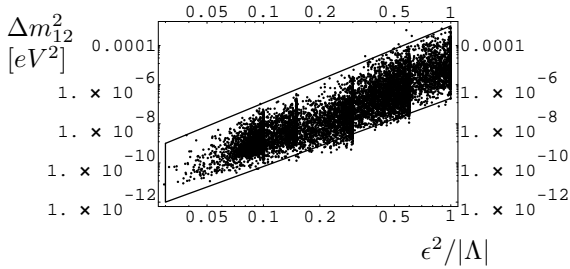}}
\end{picture}
\vspace{-20mm}

$\hskip 55mm \epsilon^2/|\Lambda|$

\vspace{-6mm} 
\caption{$\Delta m^2_{12}$ as a function of 
$\epsilon^2/|\Lambda|$. A
box is drawn to guide the eye.}
\label{fig9}
\end{figure}

%
%

\begin{figure}

\noindent
$\Delta m^2_{12}$

\noindent
$[eV^2]$

\begin{picture}(0,4)
\put(0.5,0){\includegraphics[width=70mm]{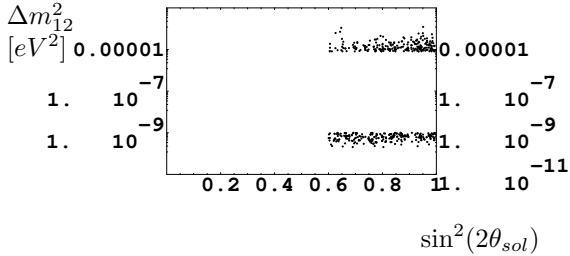}}
\end{picture}
\vspace{-20mm}

$\hskip 55mm \sin^2(2\theta_{sol})$

\vspace{-6mm} 
\caption{$\Delta m^2_{12}$ versus $\sin^2(2\theta_{sol})$ 
for those points which have $\Delta m^2_{23}$ and 
$\sin^2(2\theta_{atm})$ correct and at the same time fit either 
the vacuum or LA-MSW solutions.}
\label{fig10}
\end{figure}

\noindent
Finally we have to check if it is possible to have masses in the range
to solve the solar neutrino problem. We found~\cite{numass} that the
relevant parameter for this purpose is
\beq
(\epsilon_1^2+\epsilon_1^2+\epsilon_1^2)/|\Lambda|
\equiv
\epsilon^2/|\Lambda|
\eeq
Depending on this quantity in the range
$0.01 \le \epsilon^2/|\Lambda| \le 1$
the solar neutrino problem can be solved: Low values give {\it just-so} 
solutions, high values tend to give large angle MSW (LA-MSW). This is
illustrated in Figure~\ref{fig9} and in Figure~\ref{fig10}.

\begin{figure}
\noindent
$L_{\chi_0}$

\noindent
$(cm)$

\begin{picture}(0,7.5)
\put(0.5,0){\includegraphics[height=80mm]{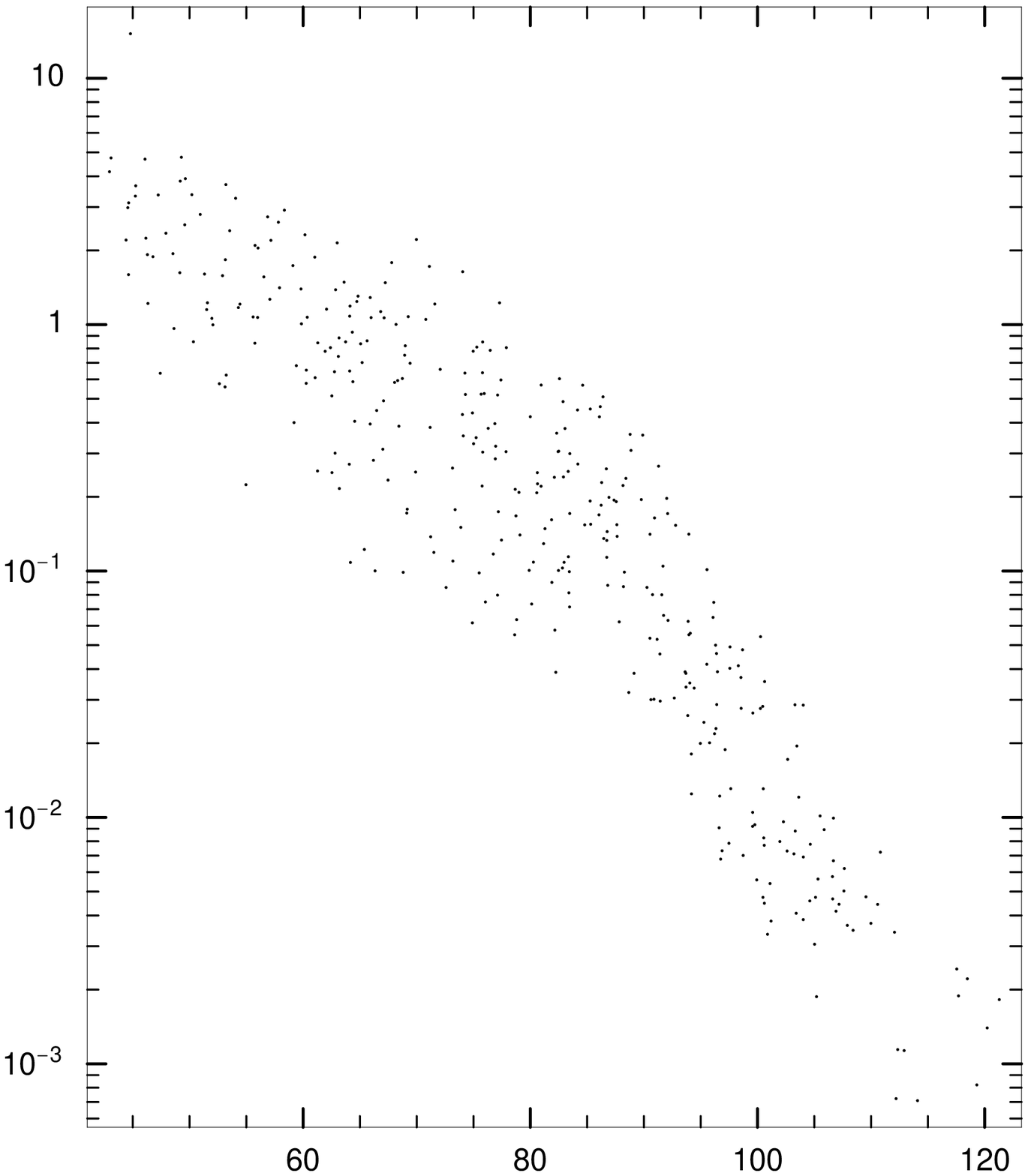}}
\end{picture}

\vskip-8mm

$\hskip 55mm m_{\chi_0}$ $(GeV)$ 

\vskip -8mm
\caption{Neutralino decay length in cm}
\label{fig12}
\end{figure}

\noindent
Another question of relevance that we addressed was the study of the
decay length of the lightest neutralino. This is important, because if
the decay length is greater than the detector, then in practice it
will be invisible like in the MSSM. As we can see in Figure~\ref{fig12},
that is not the case, the neutralino decays well inside the detector
leading to novel signatures.

\section{Conclusions}

There is a viable model for SBRP that leads to a very rich
phenomenology, both at laboratory experiments, and at present (LEP)
and future (LHC, LNC) accelerators.
We have shown that the radiative breaking of both the Gauge
Symmetry and $R_P$ can be achieved.
In these type of models neutrinos have mass and can decay thus
avoiding the critical density argument. They also can evade the BBN
limits on a $\nu_{\tau}$ on the MeV scale.
Most of these phenomenology can be described by an effective model
with bilinear explicit $R_P$ violation.
We have calculated the {\it one--loop} corrected masses
and mixings for the
neutrinos in a completely consistent way, including the RG equations
and correctly minimizing the potential.
We have shown that it is possible to get bimaximal solutions for both
the atmospheric and solar neutrino problems.
We emphasize that the 
lightest neutralino decays inside the detectors,
thus leading to a very different phenomenology than the MSSM.

\end{document}